\documentclass[aps,prl,twocolumn,groupedaddress,showpacs]{revtex4}

\usepackage{amssymb}
\usepackage{amsmath}
\usepackage{graphicx}
\usepackage{natbib}
\usepackage{nicefrac}
\usepackage{multirow}
\usepackage{comment}

\begin{document}

\title{Field Tuning the \emph{g} Factor in InAs Nanowire Double Quantum Dots}

\author{M. D. Schroer}
\author{K. D. Petersson}
\author{M. Jung}
\author{J. R. Petta}
\affiliation{Department of Physics, Princeton University, Princeton, NJ 08544, USA}

\date{\today}

\begin{abstract}
We study the effects of magnetic and electric fields on the $g$ factors of spins confined in a two-electron InAs nanowire double quantum dot.  Spin sensitive measurements are performed by monitoring the leakage current in the Pauli blockade regime.  Rotations of single spins are driven using electric-dipole spin resonance.  The $g$ factors are extracted from the spin resonance condition as a function of the magnetic field direction, allowing determination of the full g-tensor. Electric and magnetic field tuning can be used to maximize the $g$-factor difference and in some cases altogether quench the EDSR response, allowing selective single spin control.
\end{abstract}


\maketitle

The desire for a fully controllable and scalable quantum computer places several stringent conditions on its constituent qubits  \cite{Divincenzo00thephysical}.  Recent implementations of spin-qubits in GaAs heterostructures have made impressive advances, resulting in a well characterized system in which spin initialization, control, and readout have all been achieved \cite{Petta2005,Koppens2006,Bluhm2011}.  For GaAs-based qubits, rapid and selective control of single spins remains a challenge.  Single spin rotations have been performed using traditional electron spin resonance (ESR) as well as electric-dipole spin resonance (EDSR), but the highest achieved manipulation rate in GaAs is still 2 orders of magnitude slower than the exchange gate \cite{Koppens2006,Golovach2006,Nowack2007,Petta2005}. Encoding the qubit in two- or three-electron spin states can eliminate the need for single spin control, but requires local magnetic field gradients or exquisite control of the exchange interaction \cite{Levy2002,DiVincenzo2000}.

``Spin-orbit qubits'' based on InAs nanowires have recently been proposed as an alternative to the GaAs lateral quantum dot system \cite{Flindt2006,Nadj-Perge2010}.  In the spin-orbit qubit, the qubits are dressed states of spin and orbital degrees of freedom, due to the strong spin-orbit coupling of InAs. Electrical control of the qubit's orbital component allows Rabi frequencies on the order of 60 MHz to be achieved \cite{Golovach2006,Nadj-Perge2010}.  While strong spin-orbit coupling enables fast spin rotations, it has the potential to introduce several complications. In particular, the electronic $g$ factors, spin relaxation time, and EDSR rotation rates are expected to vary with magnetic field direction \cite{Takahashi2010}.  Moreover, the spin-orbit interaction couples the spin to the orbital component of the wavefunction, leading to a g-tensor that is sensitive to the gate-tunable confinement potential.

In this Letter, we explore the effects of strong spin-orbit coupling on spins confined to an InAs nanowire double quantum dot.  We demonstrate magnetic and electric field control of the $g$ factors for each quantum dot. The EDSR spin manipulation rate is a sensitive function of magnetic field direction, allowing the Rabi frequency to be maximized. For specific magnetic field directions the EDSR response can be dramatically reduced.  Our results show that electric and magnetic field tuning of the $g$ factor can be used to optimize the EDSR driving rate and allow for selective single spin control.

Our device, shown in Fig.\ 1(a), is fabricated on a high resistivity, oxidized silicon substrate. Using electron beam lithography, we define an array of gate patterns consisting of a series of 20 nm thick Ti/Au gates, spaced at a 60 nm pitch. Two large side-gates allow the transparency of the nanowire leads to be tuned. A 20 nm layer of SiN$_x$ is then deposited as a gate dielectric using plasma enhanced chemical vapor deposition  \cite{Nadj-Perge2010}. Single crystal InAs nanowires with a zinc blende structure are grown using a gold catalyzed vapor-liquid-solid process in a metal-organic chemical vapor deposition reactor \cite{Schroer2010}.  Nanowires are removed from the growth substrate with ultrasonication in ethanol and then dispersed on the gate array. The final fabrication step involves defining low resistance ohmic contacts to nanowires with diameters of $\sim$50 nm that have fallen across a gate pattern. The sample was measured in a dilution refrigerator equipped with a vector magnet system and high frequency coaxial wiring.

We first determine the charge stability diagram by measuring the current at finite applied bias, as shown in Fig.\ 1(b).  The absence of finite bias triangles in the lower left hand corner of the plot indicates the double quantum dot has been completely emptied of free electrons. In this region $\left(N_{L},N_{R}\right)$ = $\left(0,0\right)$, where $N_L$($N_R$) is the number of electrons in the left(right) dot. Measurements at high bias and with relatively transparent tunnel barries have confirmed our identification of the (0,0) charge state. We focus on the two-electron regime, where Pauli blockade results in current rectification at the (1,1)$\leftrightarrow$(2,0) charge transition, as first observed in vertical double quantum dots by Ono \emph{et al.} \cite{Ono2002}. At positive bias, charge transport at the (1,1)$\leftrightarrow$(2,0) transition occurs freely through a cycle of (1,0)$\rightarrow$(2,0)$\rightarrow$(1,1)$\rightarrow$(1,0), where steps with double occupancy are limited to spin singlet states due to the $\sim$9 meV exchange splitting of the (2,0) state.  Under negative bias, the cycle is reversed and the system loads the second electron into the (1,1) charge state, where the singlet and triplet states are nearly degenerate. If a (1,1) triplet state is loaded, the charge transition to (2,0) will be blocked due to Pauli exclusion. The left and right insets to Fig.\ 1(b) show the current through the double dot at positive and negative bias, exhibiting the voltage bias dependence characteristic of Pauli blockade.

\begin{figure}[t]
\begin{center}
		\includegraphics[width=\columnwidth]{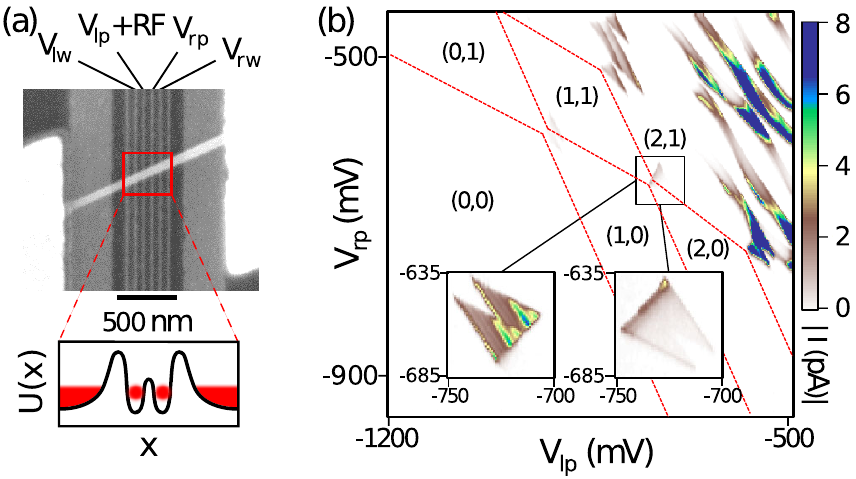}
\caption{\label{fig1} (Color online) (a) Scanning electron microscope image illustrating the device geometry. Four gates are sufficient to fully define the double quantum dot with this device, resulting in a double well confinement potential. The double dot's coupling to the right and left leads are controlled with voltages $V_{rw}$ and $V_{lw}$ respectively, while $V_{rp}$ and $V_{lp}$ control the interdot coupling and occupation. (b) Current measured through the device as a function of gate voltages $V_{rp}$ and $V_{lp}$, with source drain bias $V_{sd}$ = -4 mV.  Dashed lines are superimposed to illustrate the double dot charge stability diagram. Left inset:  $(1,1)\leftrightarrow (2,0)$ transition at  $V_{sd}$ = +4 mV. The apparent excited state is a resonance associated with the one-dimensional leads that contact the double quantum dot \cite{Mottonen2010}. Right inset: $(1,1)\leftrightarrow(2,0)$ transition at $V_{sd}$ = -4 mV. The current in the insets has been multiplied by a factor of two.}
\end{center}	
\vspace{-0.6cm}
\end{figure}

In the absence of spin-flips or cotunneling, the leakage current in the Pauli blockade regime will be zero.  Processes which drive rotations from the (1,1) triplet states to the (1,1) singlet state will lift Pauli blockade, resulting in a measurable leakage current.  Pauli blockade enables spin-dependent readout and has been used to detect the mixing of spin states due to hyperfine fields, ESR, and EDSR in GaAs and InAs quantum dots \cite{Koppens2005,Koppens2006,Nowack2007,Nadj-Perge2010}.

We measure the leakage current in the Pauli blockade regime as a function of magnetic field $B$ and the frequency $f$ of the ac excitation applied to a depletion gate (see Fig.\ 2).  A zero field peak is evident, corresponding to rapid singlet-triplet mixing due to the hyperfine fields \cite{Koppens2005,Johnson2005a}.  We extract a hyperfine field $B_{N}= 3.3$ mT from these data using the theory of Jouravlev \textit{et al.} \cite{Jouravlev2006}. Previous reports indicate $B_{N}$ = 0.66 and 1.5 mT in InAs quantum dots, with a value that depends on the number of nuclei the electron spin has appreciable overlap with \cite{Pfund2007,Nadj-Perge2010}.   Away from zero field, the leakage current is non-zero only when the EDSR resonance condition is satisfied.  The data in Fig.\ 2(a) are measured for $\mathbf{B}\: ||\: \hat{z}$, while the data in Fig.\ 2(b) are measured for $\mathbf{B}\: ||\: \left(\hat{x}+\hat{z}\right)$.  Vertical cuts through the data, shown in Fig.\ 2(d), clearly indicate that the $g$ factor of each electron is sensitive to the magnetic field direction, with well-resolved EDSR peaks visible for each spin in the lower trace.

\begin{figure}[t]
\begin{center}
		\includegraphics[width=\columnwidth]{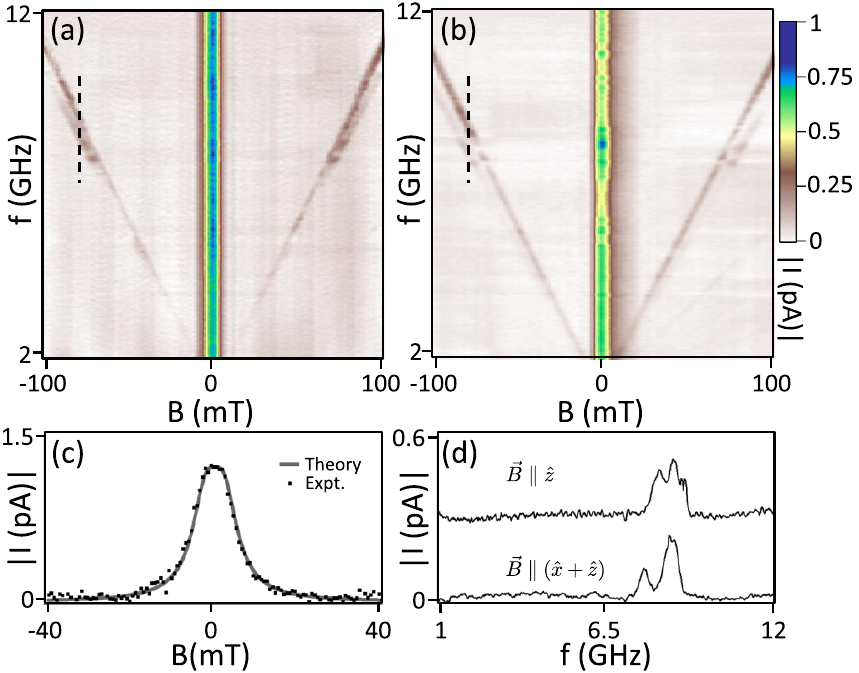}
\caption{\label{fig2} (Color online) Leakage current measured as a function of magnetic field and applied frequency for (a) $\mathbf{B}\: ||\: \hat{z}$ and (b) $\mathbf{B}\: ||\: \left(\hat{x}+\hat{z}\right)$. The zero field peak, shown in (c), is due to hyperfine field mixing of the spin states, with a fit to the data yielding $B_{N} = 3.3$  mT. The spin resonance condition is satisfied when $ hf = g \mu_{B}|\mathbf{B}|$, resulting in a peak in the leakage current. Cuts through the data in (a) and (b) are shown in (d), demonstrating the dependence of the $g$ factors on magnetic field direction. The upper trace is offset by 0.3 pA for clarity.}
\end{center}	
\vspace{-0.5cm}
\end{figure}

We quantitatively measure the directional dependence of the $g$ factors by fixing the magnetic field magnitude at $|\mathbf{B}|$ = 80 mT and rotating the field direction.  Data are shown in Figs.\ 3(a)--(c) for three different field rotations.   In each case the position and intensity of the EDSR resonances evolve with the field direction.   For each magnetic field direction we fit each resonance peak to a Gaussian and extract the $g$ factor. Fitting is only performed for regions of the data where the resonance is clearly visible.  After the $g$ factors have been extracted, we perform a simultaneous curve fit of the data sets in Fig.\ 3(a--c) to a general model of a 3D anisotropic $g$ factor:

\begin{eqnarray}
g\left(\mathbf{ B}\right) = \frac{\sqrt{g_{1}^2 B_{1}^2 + g_{2}^2 B_{2}^2+g_{3}^2 B_{3}^2}}{\left|\mathbf{B}\right|}
\end{eqnarray}

Here $g_1$, $g_2$, and $g_3$ are the values of the $g$ factors for the three principal axes, and $B_1$, $B_2$ and $B_3$ are the magnetic field components along these axes \cite{Petta2002,Brouwer2000}. Our fitting also takes into account field offsets of 2--3 mT due to trapped flux in the superconducting magnet coils.

\begin{figure*}
\begin{center}
		\includegraphics[width=2\columnwidth]{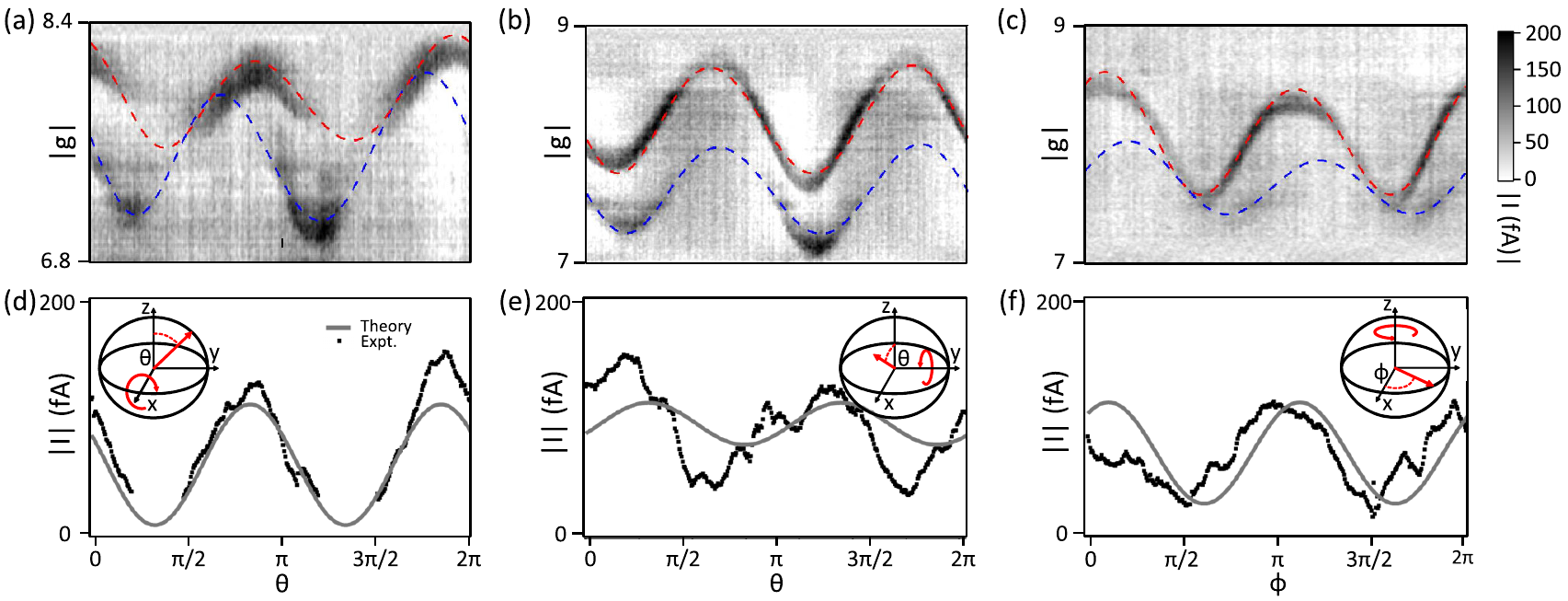}
\caption{\label{fig3} (Color online) (a--c) Leakage current measured as a function of magnetic field direction and frequency for the rotations depicted in the insets below.  The y-axes have been rescaled from frequency to units of $g$, taking a fixed magnetic field amplitude of 80 mT; all angles are radians.  In this coordinate system, the unit vector describing the nanowire axis is approximately $(\theta,\phi) = (1.2,3.1)$. A fit to Eqn.\ 1 has been superimposed on the data. (d--e) Pumped current extracted from the upper resonances in (a--c), along with a fit to Eq. (3).}
\end{center}
\vspace{-0.6cm}
\end{figure*}

The best fit values for the $g$ tensor are listed in Table I.  The principal axes' $g$ factors vary from approximately 7 to 9, suppressed from the bulk value of $|g|$ = 14.7 through a combination of spin-orbit coupling and confinement \cite{Kiselev1998,Csonka2008,Nilsson2009a}.  In each case, the $g$ tensor contains one large axis, and two approximately equal and smaller axes, appearing to mimic the cylindrical symmetry of the nanowire itself.  However, we find that the large axis is oriented in different directions for each electron, and in general is not aligned with the nanowire axis. These results suggest that the orientation of the $g$ tensor is largely determined by the anisotropy of the confinement potential, not the underlying one-dimensional symmetry of the nanowire.

We performed a second set of measurements with the device tuned to a very different gate voltage configuration in order to measure the sensitivity of the $g$ tensor to changes in the confinement potential. The data in Fig.\ 3 were acquired with the double quantum dot tuned such that the tunnel rates to the left and the right leads were balanced. To explore a different confinement potential, we unbalanced the tunnel rates by varying the voltages on the local electrodes, thereby tuning the double quantum dot into a regime where transport shows a large, asymmetric cotunneling peak [see Fig.\ 4(a)]. Figure 4(b)--4(d) map out the $g$ factor as a function of field orientation for the `unbalanced' case. The $g$ tensor has been dramatically altered, with extracted values for the principal axes directions and $g$ factors summarized in Table I. For this gate voltage configuration, the upper resonance is highly anisotropic and shifted to higher frequencies. The lower resonance is broadened significantly, which we attribute to enhanced dynamic nuclear polarization \cite{Vink2009}.

We now focus on the field dependence of the EDSR amplitude, which is expected to depend on both the magnitude and direction of the magnetic field \cite{Golovach2006}. Figure\ 3(d)--3(f) show the current extracted from the upper resonance of Fig.\ 3(a)--3(c) respectively.  We find that the on resonance current is strongly modulated by field orientation.  Golovach \emph{et~al.} \cite{Golovach2006} have shown that the effective magnetic field for spin-orbit driven EDSR in a two-dimensional system is given by

\begin{eqnarray}
	\mathbf{B}_{so}(t) = 2 \mathbf{B} \times \left(\mathbf{\Omega_0}\sin{\omega_{ac}t}\right)
\end{eqnarray}

\noindent

\begin{figure}[b]
\begin{center}
		\includegraphics[width=\columnwidth]{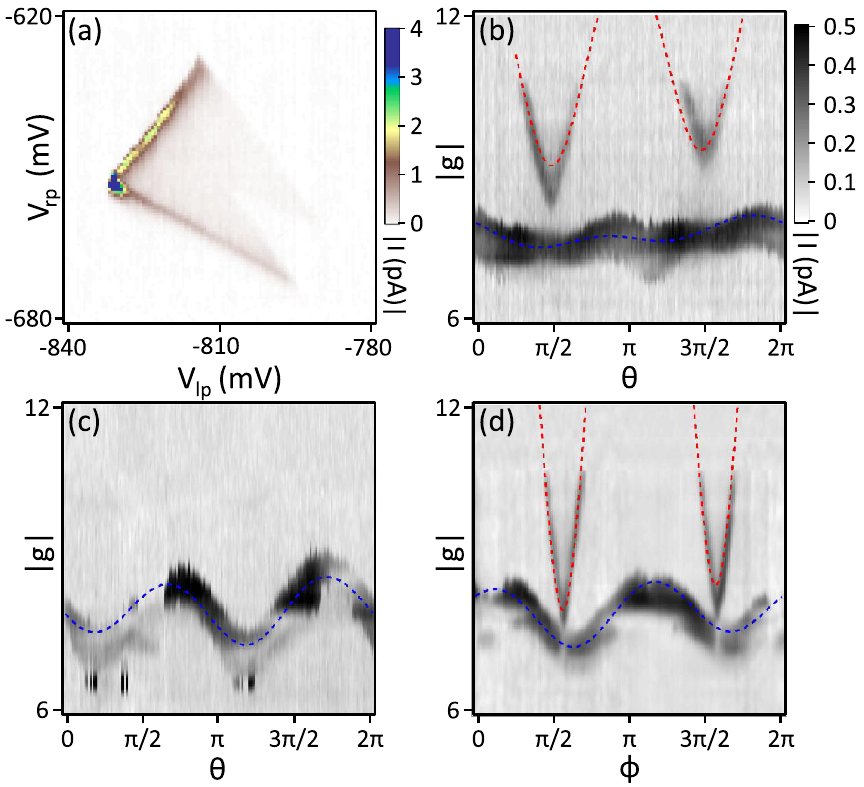}
\caption{\label{fig4} (Color online) (a) Measured current through the double dot as a function of $V_{lp}$ and $V_{rp}$ for $V_{sd}$ = -4 mV in the case of unbalanced tuning. (b)--(d) Evolution of the resonance condition for magnetic field rotations about the $x$, $y$ and $z$ axes respectively.  In (b)--(d) the $y$ axes have been rescaled into units of $g$, taking a fixed magnetic field amplitude of 80 mT.}
\end{center}
\vspace{-0.6cm}
\end{figure}

 \noindent where $\mathbf{\Omega_0}$ is a dimensionless ``spin-orbit vector" determined by the applied electric field and the local spin-orbit parameters. In the absence of an applicable theory, we generalize the concept of the spin-orbit vector to three dimensions. While Eqn.\ 2 gives the Rabi frequency, $f_{r} = g \mu_B B_{so}/2 h$ for on resonance driving, the dependence of the pumped current is less clear.  In general, the current will not be proportional to $f_{r}$ as the fluctuating nuclear field shifts the applied excitation on and off of resonance.  We have studied the induced current using a simple rate equation model, similar to the approach used by Koppens \emph{et~al}. \cite{Koppens2007}. For $B_{so}<B_{N}$ we find

\vspace{-0.2 cm}
\begin{eqnarray}
	I_{EDSR} = \frac{e \Gamma_{i} |\mathbf{B_{so}}|^2}{2 {B_{N}}^2}
\end{eqnarray}
\vspace{-0.2 cm}

\noindent
where $\Gamma_{i}$ is the interdot tunneling rate.  Fits to Eqn.\ 3 are displayed in Fig.\ 3(d)-3(f).  Equation 3 reproduces the periodic modulation of the pumped current; however we are not able to simultaneously fit the three individual rotation sweeps.  The discrepancy from the behavior predicted by Eqs.\ 2 and 3 is not well understood but may involve corrections due to the magnetic field dependence of spin-orbit matrix elements \cite{Takahashi2010}.

\begin{table}[t]
\begin{center}
	\vspace{-0.5 cm}
	\caption{$G$-tensor Parameters.  Values are given for both the balanced case of Fig.\ 3, and the `unbalanced' case of Fig.\ 4.  $\alpha$, $\beta$ and $\gamma$ are Euler angles (in radians) which define the transformation from the measurement coordinate system to the $g$ tensor coordinate system, using the $ZX^\prime Z ^{\prime\prime}$ convention.}
	\begin{tabular}{c c c c c c c c}
		\hline
		\hline
		& Dot & $|g_1|$ & $|g_2|$ & $|g_3|$ & $\alpha$ & $\beta$ &$\gamma$\\ \hline
		\multirow{2}{*}{Balanced}&A & 9.1 & 7.8 & 7.5 & 1.9 & 2.1 & -0.25\\
		& B & 8.4 & 7.3 & 7.0 & -0.81 & 1.0 & 1.5\\ \hline
		\multirow{2}{*}{Unbalanced}&A & 22 & 12 & 8.0 & 1.8 & 1.9 &-0.21\\
		&B & 8.8 & 7.6 & 7.4 &-1.2 & 1.0 & 0.73\\ \hline
\end{tabular}
\vspace{-0.8cm}
\end{center}
\end{table}

Equation 3 may also be used to estimate the magnitude of the spin-orbit field.  With a slightly different tuning, we have observed EDSR currents exceeding 1.5 pA at $|B|=150$ mT.  For this tuning, we measure positive-bias (nonblockaded) currents of $\sim$35 pA. Assuming the interdot tunnel coupling is rate limiting, this results in a lower bound estimate of $\Gamma_{i} \sim$ 220 MHz, leading to a spin-orbit field $|\mathbf{B_{so}}|=$ 1.0 mT.  Finally, we may estimate the spin-orbit length $l_{so}$ as
\vspace{-0.1cm}
\begin{eqnarray}
	l_{so} = \frac{B}{B_{so}} \frac{2 \hbar^2 e |E|}{m_e \Delta^2}
\end{eqnarray}
where $|E|$ is the applied electric field, $m_e$ is the effective mass, and $\Delta$ is the level spacing \cite{Golovach2006}.  We estimate $|E|$ as 1 mV/60 nm, and find $\Delta$ = 9 meV from finite bias measurements.  With these parameters, we find $l_{so} \sim 170$ nm.  Previous studies have reported values for the spin-orbit length in InAs nanowires of 100--200 nm  \cite{Fasth2007,Dhara2009}.

The large variation of the $g$ factors with electric field tuning suggests that $g$ tensor modulation might play a significant role in electrically driven spin manipulation \cite{Kato2003}. Assuming the change in the $g$ factor is purely linear in electric field, we find a maximum $\partial$g/$\partial V_g$ $\sim$ 0.03/mV. Taking $g$ = 8 and assuming an applied magnetic field of 80 mT, this yields an effective ac magnetic field of 0.3 mT, which is of the same order as the calculated spin-orbit field. Our results suggest that $g$ tensor modulation may be significant for specific electric and magnetic field configurations.

We have demonstrated magnetic and electric field tuning of the electron $g$ factors in an InAs nanowire double quantum dot, which have important implications for spin-orbit qubits. The ability to selectively control single spins may be obtained by tuning magnetic and electric fields to (1) maximize the difference between the $g$ factors of neighboring spins and (2) maximize the difference in the response of neighboring spins.  A linear array of quantum dots in an InAs nanowire could be used to implement a Loss-DiVincenzo style spin register, where neighboring spins are tuned out of resonance with electric field tuning of the $g$ tensor.  The combination of spectral selectivity with the localized nature of the on-chip generated microwave fields may allow the development of scalable quantum dot arrays in InAs nanowires.

\begin{acknowledgments}
We thank S. Frolov, C. Kloeffel, and D. Loss for helpful discussions. Research at Princeton was supported by Army Research Office Grant No.\ W911NF-08-1-0189, DARPA QuEST Grant No. HR0011-09-1-0007, and the NSF through the Princeton Center for Complex Materials and the Imaging Analysis Center (DMR-0819860). Research was carried out in part at the Center for Functional Nanomaterials, Brookhaven National Laboratory, which is supported by DOE BES Contract No.\ DE-AC02-98CH10886.
\end{acknowledgments}
\vspace{-0.55cm}

\bibliographystyle{apsrev}
%\bibliography{anisotropy}

\end{document}